\documentstyle[preprint,aps,floats,epsfig]{revtex}

\def\real{I\negthinspace R}

\def\half{\textstyle{1\over2}}

\def\ie{{\it i.e.,}}

\newcommand{\be}{\begin{equation}}
\newcommand{\ee}{\end{equation}}
\newcommand{\bea}{\begin{eqnarray}}
\newcommand{\eea}{\end{eqnarray}}
\newcommand{\bml}{\begin{mathletters}}
\newcommand{\eml}{\end{mathletters}}

\begin{document}
\preprint{DTP/02/41, hep-th/0208037}
\draft
\tighten

\title{Spacetime structure of the global vortex}
\author{Ruth Gregory$^1$, Caroline Santos$^2$}
\address{~$^1$ Centre for Particle Theory, 
Durham University, South Road, Durham, DH1 3LE, U.K.\\
$^2$ Departamento de F\'\i sica da Universidade do Porto \&
%, Rua do Campo Alegre 687, 4150-Porto, Portugal\\
%$^3$ 
Centro de F\'\i sica da Universidade do
Porto, Rua do Campo Alegre 687, 4150-Porto, Portugal.}
\date{\today}
\setlength{\footnotesep}{0.5\footnotesep}
\maketitle

\begin{abstract}
We analyse the spacetime structure of the global vortex and its maximal 
analytic extension in an arbitrary number of spacetime dimensions. 
We find that the vortex compactifies space on the scale of the Hubble 
expansion of its worldvolume, in a manner reminiscent of that of the 
domain wall. We calculate the effective volume of this compactification 
and remark on its relevance to hierarchy resolution with extra dimensions. 
We also consider strongly gravitating vortices and derive bounds on the 
existence of a global vortex solution.
\end{abstract}

\pacs{PACS numbers: 04.40.-b, 04.50.+h, 
11.27.+d \hfill hep-th/0208037}

\section{Introduction}

The global vortex is a topologically non-trivial
vacuum defect solution in $2+1$ dimensions (or $2+1+p$ dimensions,
in which case the vortex is a $p$-brane).
Derrick's theorem would normally lead us to expect that there
could be no such soliton, however, the vortex evades Derrick's
theorem by having not a finite energy, but a logarithmically 
divergent one.
This divergence naturally means that when the gravitational
interactions of the vortex are included, the story becomes rather
interesting.
At first, it was assumed that the spacetime of a global vortex
would be static, like its local Nielsen-Olesen cousin, however, 
such an assumption led to an exact, but singular metric 
\cite{CK1} outside the core of the vortex. 
It was rapidly realised that any static global vortex (including those 
derived from a sigma model) would be singular
\cite{RG1,GOR}.

The global vortex is however, one of a family of global defect
solutions. The domain wall, which separates regions of discrete
vacua, is perhaps the simplest, however, there is also the global
monopole \cite{BV} which has a linearly divergent energy 
and is unstable \cite{GMU}.
Of the three global defects (in four dimensions) only the wall
has a finite energy, indeed in more than four dimensions where
there are further global $p$-brane solutions analogous to the
monopole \cite{OV}, the wall continues to be the only defect 
with a finite energy (per unit brane volume). 
Curiously this lack of finiteness of monopole energy does not
lead to gravitational singularities as it does for the vortex,
for the global monopole has a well-defined static gravitational field
\cite{BV} (as does its higher dimensional descendents \cite{OV}) 
which is an asymptotically locally flat (ALF) spacetime with a 
global conical deficit angle.

The gravitational field of the domain wall is well studied
(e.g.\ \cite{AV,IS,GWG,BCG}), and
from the perspective of an observer on the wall is non-static, having a
Hubble expansion parallel to the wall. This however is known
to be illusory, in the sense that it is the wall itself which is in
motion, in a static Minkowski spacetime \cite{GWG}. The vacuum
domain wall is in fact an accelerating bubble, which contracts in from
infinite radius, decelerating until it reaches a minimum size, 
then re-expanding out again to future null infinity. 
Constant time surfaces are then topologically
spherical, and the wall can be viewed as self-compactifying spacetime on a
scale roughly of its inverse energy. This picture of course refers to
an infinitesimally thin wall, however, the thick topological defect
solution has the same geometry, with the scalar field simply following
the contours of the bubble \cite{BCG}.

In the broader context of the global defect family, the singular nature of
the static global vortex spacetime was anomalous.
The resolution of the puzzle was of course to add time dependence
to the vortex metric \cite{RG2}, which allowed for the existence of a
non-singular solution to be demonstrated. This time dependence was of
the same nature as that of the vacuum domain wall spacetime,
\ie\ a Hubble expansion along the length of the vortex, however, the
full global spacetime structure of the global vortex and its maximal
analytic extension was not explored, and indeed there have been claims
that the cosmological event horizon present in the vortex spacetime
is unstable to perturbations and becomes singular \cite{W}.

More recently the issue of global defect solutions has acquired a
new dimension (quite literally!) as the renaissance of the braneworld
scenario has resulted in a search for new exotic compactifications.
The braneworld scenario is a compelling picture in which instead of
having a Kaluza-Klein compactification of extra dimensions,
we have relatively large extra dimensions, but are confined
to a four-dimensional submanifold within this larger spacetime.
The Randall-Sundrum (RS) scenario \cite{RS} is an example of a particular
warped compactification which has many features in common with the
more `realistic' string-motivated heterotic M-compactifications \cite{LOW}.
The RS scenario consists of one or two reflection symmetric
domain walls at the edge of an anti de Sitter bulk, with only gravity
propagating in the bulk. Naturally a wall can only ever give a
warped compactification with one extra dimension, and it is interesting
to explore warped compactifications of higher codimension.

Warped compactifications with local defects have been explored in a
variety of papers \cite{LC}, however, these compactifications are
often either asymptotically flat in the extra dimensions, or have
singularities. The situation with global defect compactifications
is slightly different. In vacuum, a static vortex compactification
is singular \cite{CK2}, although analogous to the four dimensional global
string, this singularity can be removed either by the addition of time
dependence (see the next section), or by adding a negative bulk 
cosmological constant \cite{RG3}.
Higher co-dimension global defect spacetimes based on the global
monopole were considered in \cite{OV}, and analogous to the global
monopole tend to exhibit global deficit angles.
In addition, warped stringy compactifications with Hubble expansion
on the brane have been studied in \cite{Per}.

The question we are interested in in this paper is the spacetime
structure of the global vortex. The presence of an event horizon,
as well as the Hubble expansion along the vortex, indicates that something
similar to the domain wall spacetime might in fact occur.
In \cite{DG} an extension of the vortex
spacetime was proposed which glued together a vortex and anti-vortex
(V${\bar {\rm V}}$) space to form a sphere with the vortex and anti-vortex
at antipodal points. It was pointed out in \cite{OV} that this
solution could have an alternate interpretation as a vortex around
an equatorial circle of an $S^3$, just as the higher dimensional
global $p$-branes can sit on big circles of de-Sitter spheres.
However, these two situations are somewhat different; the global $p$-brane
has an ALF spacetime, so adding a global $p$-brane to the already
compact de Sitter space is simply a question of embedding.
The vortex on the other hand has a strong effect on spacetime far from its
core, and so if it lies on any great circle of a compact sphere, then
it can only do so by actually self-compactifying spacetime.
In this paper, we show that this is indeed what happens, by considering
a generalization of the coordinate transformation used in \cite{RG2} to
explore the cosmological event horizon (CEH). We also show how to extend
the metric beyond the CEH, and derive the maximal analytic extension of
the global vortex spacetime.

\section{The global vortex phase plane}

In this section we review and extend the arguments of \cite{RG2,RG3}
for the existence of a global vortex solution in $p+3$ dimensions.
We will be looking for a topologically nontrivial solution of the
field theoretic lagrangian
\be
{\cal L} = (\nabla_\mu \Phi)^\dagger \nabla ^\mu \Phi -
{\lambda \over 4} ( \Phi^\dagger\Phi -\eta^2)^2
\ee
in an otherwise empty spacetime.  By writing
\be
\Phi = \eta X e^{i\chi}
\ee
we can reformulate the complex scalar field into two real interacting
scalar fields, one of which ($X$) is massive, the other ($\chi$) being the
massless Goldstone boson responsible for the divergent energy of the
vortex. In this way, the low energy theory is seen
to be equivalent in $n(=p+3)$ dimensions to an $(n-2)-$form potential. For
example in four dimensions, the $\chi$-field is equivalent to a
Kalb-Ramond $B_{\mu\nu}$ field, and the effective action for the
motion of a global string is the bosonic part of the superstring action.

A vortex solution is characterised by the
existence of closed loops in space for which
the phase of $\Phi$ winds around $\Phi=0$ as a closed loop
is traversed. This in turn implies that $\Phi$ itself has a zero within
that loop, and this is the core of the vortex.
From now on, we shall look for a solution describing a vortex with 
unit winding number \ie\ $\chi=\theta$, where $\partial_\theta$
is some Killing vector of spacetime with closed circular orbits.

The boost symmetry of the energy momentum tensor parallel to the defect
means that the most general metric can be written in the form
\be
ds^2 = e^{2A(r)} H^{-2} g_{\mu\nu} dx^\mu dx^\nu - dr^2
- C^2(r) d\theta^2 \label{metric}
\ee
where $\mu, \nu =0, 1...p$ are the coordinates parallel to the defect,
$g_{\mu\nu}$ is a constant unit-curvature de Sitter metric,
and $H$ is a constant to be determined which will represent
the Hubble parameter of the expansion on the vortex braneworld.

The Einstein equations for the vortex are found to be
\bml\label{geometry}\bea
\left[e^{(p+1)A}\,C^\prime\right]^\prime &
=&-\epsilon \,e^{(p+1)A} \,
\left(\frac{2X^2}{C} + \frac{C\,(X^2-1)^2}{2(p+1)}
\right)
\label{geometry1}\\
\left[C\,e^{(p+1)A} \,A^\prime\right]^\prime
-C\,H^2\,p\,e^{(p-1)A}&=
&-\epsilon \,e^{(p+1)A}
\,\left(\frac{C\,(X^2-1)^2}{2(p+1)}\right)
\label{geometry2}\\
(p+1)\,\left[\frac{p}{2}(H^2\,e^{-2A}-A^{\prime 2})
-\frac{A^\prime C^\prime}{C}\right]&=
&-\epsilon X^{\prime 2} + \frac{\epsilon X^2}{C^2}
+ \frac{\epsilon}{4} (X^2-1)^2
\label{geometry3}\\
\left[C\,e^{(p+1)A}\,X^\prime \right]^\prime &=&
C\,e^{(p+1)A}\,\left[\frac{X}{C^2}
+\frac{1}{2}X(X^2-1)\right] \label{vortexeqn}
\eea\eml
where we have taken $\lambda \eta^2 = 1$, and $\epsilon = \frac{\eta^2}{2}
(M_n)^{-(p+1)}$ which represents the gravitational strength of the vortex.

In what follows, unless stated otherwise, we will take the vortex 
to be weakly (or at least relatively weakly) gravitating, \ie\ 
$\epsilon \ll 1$; we comment on large $\epsilon$ in section \ref{sect:eps}. 
We can then see that $H$ is at least $O(\epsilon)$ and noting that
to leading order (\ref{vortexeqn}) implies
\be
\left [ X^2 - r^2 X^{\prime2} + {r^2\over4}(X^2-1)^2 \right ] '
= {r\over 2} (X^2-1)^2\; ,
\ee
we see that outside the core
\bml\bea
e^{(p+1)A} &=& 1 - {\epsilon\over2} - \epsilon \ln \left (
{r\over r_c} \right )
+ \epsilon \int_0^{r_c} r X^{\prime2} +
O(\epsilon^2) \label{coreA} \\
C' &=& 1 - \epsilon\mu + {\epsilon p\over (p+1)} - \epsilon \ln
\left ( {r\over r_c} \right )+ O(\epsilon^2) \label{coreC}
\eea\eml
where $\mu$ is the ``renormalized'' energy per unit length
(\ie\ ignoring the logarithmically divergent term)
\be
\mu = \int_0^{r_c} \left [ r X^{\prime2} + {X^2\over r}
+ {r\over4}(X^2-1)^2 \right]\; ,
\ee
and $r_c$ is an effective width of the vortex, of order unity,
outside of which $X\simeq 1$.
These linearized forms can be used outside the
core while $r \ll r_c e^{1/\epsilon}$.

In \cite{RG2} it was shown that a non-singular solution had to
asymptote an event horizon, \ie
\be
e^A \simeq H(r_H-r) \qquad,\qquad C \simeq C_0 ( 1 +
O(r_H-r)^2)
\label{asyhor}
\ee
as $r\to r_H$, the horizon radius. It is not difficult to show that this
remains the case for general $p$. A quick integration of
(\ref{geometry2}) then indicates that $H^2 = O(e^{-1/\epsilon})$,
a rough estimate which will be borne out by the more detailed
calculations later in this section.

In order to demonstrate the existence of the vortex, we therefore
need to verify the existence of a solution interpolating between
the core solution (\ref{coreA},\ref{coreC}) and the horizon.
To do this, we examine the field equations outside the core, \ie\  
we set $X \simeq 1$. Writing $\rho = \int\,e^{-A}\,dr$, and defining
\be\label{xandy}
x = p\frac{dA}{d\rho} + \frac{1}{C}\frac{dC}{d\rho} \qquad
; \qquad y = \frac{1}{C}\frac{dC}{d\rho}
\ee
these field equations can be written as a two dimensional
dynamical system
\bml\bea
\frac{dx}{d\rho}&=& \frac{x^2}{p}-\frac{(p+1)}{p}\,y^2 - p\,H^2
\label{xdot}\\
\frac{dy}{d\rho}&=& \frac{(p+1)}{p}\,x^2 - \frac{(p+1)}{p}\,y^2
-p(p+1)\,H^2 - xy
\label{ydot}
\eea\eml
with the constraint
\be\label{constraint}
{2\epsilon p e^{2A} \over (p+1)C^2} = -\left ( x^2 - y^2 - p^2
H^2 \right
)
\ee

The phase plane for this system is characterized by the
critical points
\bml\bea
P_{\pm}&=&H\, \left(\pm p , 0\right)\\
Q_{\pm}&=&H\,\left(\pm \sqrt{p(p+1)},
\pm \sqrt{\frac{p}{p+1}}\right)
\eea\eml
and the invariant hyperboloid
\be\label{hyperboloid}
x^2 - y^2 = p^2 \, H^2\,.
\ee
The critical points $P_{\pm}$ are saddle points while the
classification for $Q_{\pm}$ depends on the value of $p$: for
$p<7$,
$Q_+$/$Q_-$ is a stable/unstable focus while for
$p \geq 7$, $Q_+$ is an attractor and $Q_-$ a repeller.
Even so, the qualitative shape of the phase plane remains
approximately the same irrespective of the value of $p$,
provided that one re-scales the coordinates $x$ and
$y$. The phase space is represented in figure 1 where we
have chosen $p = 3$.

%%%%%%%%%%%%%%
% FIGURE 1
%%%%%%%%%%%%%%
\begin{figure}[ht]
\begin{center}
\epsfxsize=10cm
\epsffile{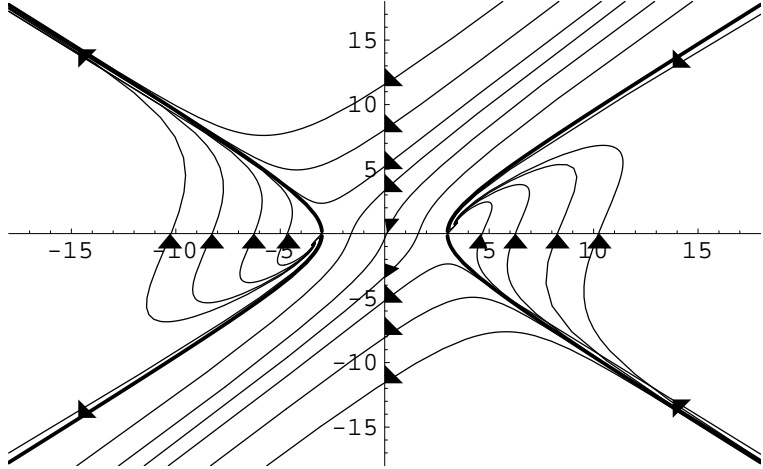}
\end{center}
\caption[phspace]{\label{phspace} The phase space $y(x)$ for
the global vortex $3-$brane. The thick line represents the
invariant hyperboloid in eqn (\ref{hyperboloid}) and the
critical points are located at $P_{\pm} = (\pm 3,0)$ and
$Q_{\pm} = \pm (2\sqrt3,\frac{\sqrt3}{2})$ where
we have fixed $H = 1$ through a redefinition of $\rho$.}
\end{figure}

The asymptotic solutions corresponding to the critical points are:
\bml\bea
P_{\pm}&:&e^{A(r)}\,\simeq \,\pm H\,(r-r_H),\;
C(r) \simeq C_0\left [ 1- {\epsilon(r-r_H)^2\over(p+2)C_0^2}
\right ], \; r \rightarrow r_H^{\pm}
\label{Ppm}\\
Q_{\pm}&:&e^{A(r)}\simeq \pm Hr \sqrt{\frac{p}{p+1}},
\; C(r) \simeq \pm r\sqrt{2\epsilon\over p+1} , \;
\qquad r \rightarrow \pm \infty \label{Qpm}
\eea\eml
where $r_H$ is the horizon of the defect in the transversal
section of the brane. Note however that the solution for $Q_\pm$
should be treated with some caution, as this lies outside the
physically allowed region of the phase plane as defined by
positivity of the constraint (\ref{constraint}). However, we will
see in the next section that $Q_\pm$ is relevant for the
analytically continued vortex spacetime outside the CEH.

Now consider the critical point $P_-$. From the constraint
(\ref{constraint}) we see that this corresponds to
the vortex horizon as $\frac{dC}{d\rho} = e^{2A} = 0$
and $\frac{dA}{d\rho} = - H$. Therefore a non-singular
solution exists for the vortex $p$-brane if a trajectory
which corresponds to the initial conditions emerging from
the core at $r_c$ approaches this critical point. 
Note that $H$ is as yet undetermined. From (\ref{xdot})
and (\ref{ydot}) we see that rescaling $H$ simply rescales the
phase plane.  We therefore determine $H$ by the requirement
that the nonsingular trajectory must pass through the initial $(x_c,y_c)$
determined from the core solution (\ref{coreA},\ref{coreC}).
This is what we now investigate. We separate the problem into two parts:
the behaviour of the trajectories from the core
to the $y$-axis, and then from the $y$-axis to the critical point.

First however, we make some qualitative remarks about
the phase plane trajectories. Using the constraint
(\ref{constraint}) we can rewrite
the dynamical system (\ref{xdot})-(\ref{ydot}) as
\bml\bea
\frac{dx}{d\rho} & = & -\frac{2\epsilon\,e^{2A}}{C^2\,(p+1)}
-y^2\\
\frac{dy}{d\rho} & = & -\frac{2\epsilon\,e^{2A}}{C^2} - x\,y
\eea\eml
which shows that $x$ is monotonically decreasing along
a trajectory, whereas $y$ is decreasing only
when $y$ and $x$ are both positive. Thus the
trajectories coming from the core (which have
$y\simeq x = O(1)$) cross the $y-$axis.  Once
they have crossed to $x<0$, either $y(x=0)$ is sufficiently
large that the trajectories
have a turning point and pass to increasing $y$,
or they hit $y=0$. The critical nonsingular trajectory 
of course asymptotes $y=0$ at $P_-$, however the others
all pass to $x,y \to -\infty$. This shows why it is important 
to analyse the crossing of the $y$-axis.

In order to characterize the trajectories near the core
we use (\ref{coreA}) and (\ref{coreC}) to see that
\be
A' \simeq {dA\over d\rho} = {-\epsilon\over (p+1)r_c} + O(\epsilon^2)
\ee
and hence for trajectories near the core
\be\label{yoverxcore}
{y\over x} = \left ( 1+{pA'C\over C'} \right )^{-1}
= 1 + {p\epsilon\over(p+1)} + O(\epsilon^2)
\ee
with
\be\label{ycore}
y_c  =  \frac{1}{r_c}
\ee
from (\ref{constraint}). These give the  initial conditions
for the integration of the dynamical system at the `edge'
of the vortex, $r_c$.  At next to leading order one gets
\be\label{leadingy}
y  =  x \left[1+\frac{p\epsilon}{(p+1)\left(1
+\epsilon\ln (x/x_c)\right)}\right]\,.
\ee
Indeed, while $x,y \gg pH$, we may ignore the $H^2$
terms in (\ref{xdot},\ref{ydot}) which reduces our system
to the Cohen-Kaplan (CK) analysis. Hence, writing $u = 
u_c - \int_{r_c}^r C dr$ (so that $u=0$ corresponds to the
CK singularity and $u=u_c \simeq 1/\epsilon$ the edge of
the vortex -- see \cite{CK2}) our solution in this r\'egime
is
\bml\bea
e^{(p+1)A} &=& {u\over u_c} \\
C^2 &=& \gamma^2 \left ( {u\over u_c}\right )^{p\over p+1}
\exp \left \{ \epsilon(u_c^2-u^2)\right \}
\eea\eml
where $\gamma = r_c + O(\epsilon)$ is of order unity and given 
by integrating out from the core.
From these we may read off:
\bml\bea
x_{CK} &=& {e^A\over C} \left [ \epsilon u
- {p\over 2(p+1)u} \right ] \\
y_{CK} &=& {e^A\over C} \left [ \epsilon u
+ {p\over 2(p+1)u} \right ] 
\eea\eml
Therefore the CK trajectory crosses the $y$-axis at
\be
y_{CK} = {2\over\gamma^2} e^{-1/2\epsilon}
e^{p\over4(p+1)} \left ( { p\epsilon\over2(p+1)}
\right )^{3p+4\over4(p+1)}
\ee
Provided that this is greater than, or even roughly of the
same order as $pH$, this will be a good approximation to
the actual nonsingular trajectory crossing the $y$-axis.

Finally, note that for the actual phase plane trajectories
\be
{dy\over dx} = (p+1) - {py\left ( (p+1)y-x\right)\over
(p+1)y^2 + p^2H^2 - x^2} > (p+1) - {py\left ( (p+1)y-x\right)\over
(p+1)y^2  - x^2}  = {dy_{CK}\over dx_{CK}}
\ee
for $y>x>0$. This means that the true trajectories are slightly
steeper than the CK trajectories, but as already noted, the
deviation is minor while $y\geq O(pH)$.

Now let us analyse the behaviour of the trajectory near $P_-$.
In the vicinity this non-degenerate critical point,
the dynamical system is approximately linear and given by
\bml\bea
\left[\begin{array}{ccc}
&\frac{d\bar x}{d\rho}&\\
&\,&\\
&\frac{d\bar y}{d\rho}&
\end{array}\right]\, =\,H {\bf M}\,
\left[\begin{array}{ccc}
&\bar x&\\
&\,&\\
&\bar y &
\end{array}\right]
\eea\eml
with
\bml\bea
{\bf M} =
\left[\begin{array}{ccc}
-2
&\,\,\,\,\,\,\,\,&
0\\
&\,\,\,\,\,\,\,\,&\\
-2(p+1)
&\,\,\,\,\,\,\,&
p
\end{array}\right]
\eea\eml
and where $\bar x = x + pH$ and $\bar y = y$, 
with $|\bar x| ,|\bar y| \ll pH$.
Calculating the eigenvectors
relative to the eigenvalues of the matrix ${\bf M}$ one 
finds that the nonsingular trajectory approaching $P_-$ is
\be\label{approaching}
\bar y = 2 \left(\frac{p+1}{p+2}\right)\,\bar x
\ee
whereas trajectories with $\bar x = 0$ repel from that point.
Note that (\ref{approaching}) leads to the asymptotic
solutions given in (\ref{Ppm}).

We would now like to estimate where this nonsingular
trajectory crosses the $y$-axis. Consider the straight line
\be
y = m(x+pH) \label{straight}
\ee
for $x>-pH$. Then, along this line we may use 
(\ref{xdot},\ref{ydot}) to obtain:
\be
{dy\over d\rho} - m {dx\over d\rho} 
= {y\over m^2p} \left [ (m^2-1)(m-1)(p+1)y
+ \left ( m(p+2)-2(p+1)\right) mpH \right ]
\ee
Thus (remembering that both $y$ and $x$ are \emph{decreasing}
along trajectories) we have:
\be
{dy\over dx}
\cases{ \leq m & for $m\geq {2(p+1)\over(p+2)}$\cr
>m & for $m=1$.\cr}
\ee
This means that the trajectory passing through 
$\left ( 0, {2(p+1)\over(p+2)} pH\right )$ misses $P_-$,
turns around and asymptotes a singular CK trajectory in the
upper left hand quadrant of the phase plane. On the other
hand, the trajectory passing through $(0,pH)$ also misses
$P_-$, but crosses $x=0$ passing to the lower left hand
quadrant (which is also a singularity). Therefore, the nonsingular
trajectory crosses the $y$ axis in the range
$\left ( pH, {2(p+1)\over(p+2)}pH \right )$.
Combining this with the fact that the true trajectories are slightly
steeper than the CK trajectories, means that we get a very good
approximation to $H$ by matching the CK trajectory at $x=0$ to
the critical trajectory (\ref{approaching}):
\be \label{Hdef}
H= h(p) \epsilon^{3p+4\over4(p+1)} e^{-1/2\epsilon}
\ee
where
\be \label{hdef}
h(p) = {2(p+2)\over \gamma^2p^2} e^{p\over4(p+1)}
\left ( {p\over2(p+1)}\right )^{7p+8\over4(p+1)}
\ee
is of order $O(1/p)$.

Therefore, we have demonstrated the existence of a nonsingular
trajectory corresponding to the exterior spacetime of a global vortex
$p$-brane with a de-Sitter like expansion parallel to the brane
with the Hubble constant of this expansion being very finely tuned
to the gravitational strength of the vortex, and given by (\ref{Hdef}).

\section{Global structure and maximal extension}

In the previous section we demonstrated the existence of a
non-singular
solution to the vortex equations, and derived a good
approximation
to the spacetime within the horizon. This spacetime has the
form of
(\ref{metric}) with a good approximation to the functions $A$
and $C$
being
given by:
\bml\bea
e^A &=& {\cases{ (1-\epsilon\ln r)^{1\over p+1} & $r<r_1$ \cr
H(r_H-r) & $r>r_1$\cr }}\\
C &=& {\cases{(1-\epsilon\ln r)^{-p\over2(p+1)} \
r^{1-{\epsilon\over2}\ln
r}
& $r<r_1$ \cr
C_0 \left [ 1 - {\epsilon \over(p+2)C_0^2} (r_H-r)^2 \right ]
& $r>r_1$\cr }}
\eea\eml
where $(1 - \epsilon \ln r_1) \simeq \epsilon^{1/2}$, 
with $r_H \simeq r_1 + \epsilon^{1/2(p+1)} H^{-1}$ and $C_0
\simeq \epsilon^{1/2} H^{-1}$.

Before deriving the maximal extension and global spacetime structure
of the vortex $p$-brane, it is first instructive to examine de Sitter
spacetime in this cylindrical coordinate system. Recall that
$n$-dimensional de Sitter spacetime can be represented as a hyperboloid in
$(n+1=p+4)-$dimensional flat Minkowski spacetime:
\be
{\bf X}_{p+1}^2 - T^2 + Y^2 + Z^2 = H^{-2}
\ee
where $H$ is the Hubble constant for the de Sitter
space. Conventionally,
one coordinatises the hyperboloid by setting $T=\sinh t$ {\it
etc}., however,
to make contact with the form of (\ref{metric}) instead
consider the transformation
\bml\bea
\left ( T, {\bf X}_{p+1}\right ) &=& H^{-1} \cos Hr \
\left ( \sinh t,\ \cosh t \ {\bf n}_{p+1} \right ) \\
Y + iZ &=& H^{-1} \sin Hr \ e^{i\theta}
\eea\eml
(where ${\bf n}_{p+1}$ is the unit vector in $(p+1)$
dimensions).
This gives the metric
\be\label{deSmet}
ds^2 = H^{-2} \cos^2Hr \left [ dt^2 - \cosh^2t\ d\Omega_p^2
\right ] -dr^2
- H^{-2} \sin^2Hr\  d\theta^2
\ee
Note how this metric satisfies the same core boundary
conditions on $A$
and
$C$ as the vortex metric, and as $r\to\pi/2H$, we have similar
behaviour
as the vortex horizon, with $e^A \sim H(r_H-r)$, although here
$C_0=H^{-1}$
rather than the $\epsilon^{1/2} H^{-1}$ of the vortex horizon.

Clearly this coordinate system does not cover the full
hyperboloid
as it does not allow $Y^2 + Z^2 > H^{-2}$. For this region we
must
instead take
\bml\bea
\left ( T, {\bf X}_{p+1}\right ) &=& H^{-1} \sinh H\tau \
\left (\cosh \xi, \ \sinh \xi \ {\bf n}_{p+1} \right ) \\
Y +iZ &=& H^{-1} \cosh H\tau \ e^{i\theta}
\eea\eml
giving
\be
ds^2 = d\tau^2 - H^{-2} \sinh^2 H\tau \left [ d\xi^2 + \sinh^2
\xi
d\Omega_p^2
\right] - H^{-2} \cosh^2 H\tau\ d\theta^2
\ee
as the metric in the region `outside' the event horizon. We therefore
see that the horizon is, as expected, simply an artifact of the choice
of coordinates, which, by splitting off the $Y$ and $Z$ directions, take
a `strip' of the hyperboloid only and the actual de Sitter spacetime is 
of course completely regular across the horizon.

Let us now turn to the vortex metric, which, in the neighborhood of
the horizon, can be easily shown to be $(p+2)$-dimensional Minkowski
spacetime times a (rather large) circle.  (Take $T=(r_H-r) \sinh t$ and
${\bf X}_{p+1}=(r_H-r) \cosh t \ {\bf n}_{p+1}$.)
We have already remarked on the similarities between this horizon
and the horizon of the de Sitter spacetime in cylindrical coordinates,
therefore we set
\bml\bea
\left ( T, {\bf X}_{p+1}\right ) &=& H^{-1} e^A
\left ( \sinh t,\ \cosh t \ {\bf n}_{p+1} \right ) \label{ring} \\
Y +iZ &=& C(r) e^{i\theta}
\eea\eml
which satisfies
\be
{\bf X}_{p+1}^2 - T^2 + Y^2 + Z^2 = C^2 + H^{-2} e^{2A} \ .
\ee
For our solution, this RHS is very nearly constant:
$C^2 + H^{-2} e^{2A} \simeq H^{-2} (1+O(\epsilon))$, and in fact even
for the rather large value of $\epsilon = 0.1$, changes by
only 1 part in a billion. This means that topologically, within its
event horizon at least, the vortex spacetime looks to be the same
as de Sitter with spherical spatial topology. Metrically, the
induced metric on the vortex hyperboloid is
\be
ds^2_{\rm ind} = H^{-2} e^{2A} \left [ dt^2 - \cosh^2 t \
d\Omega_p^2 \right ]
- C^2 d\theta^2 -  \left ( A^{\prime2} e^{2A} H^{-2} +
C^{\prime2} \right)
dr^2
\ee
Because of the presence of $H^{-2}$, this appears at first sight
to be different to (\ref{metric}), but in fact this term is only
significantly different from $1$ in the CK r\'egime, $r<r_1$. Here
$C' \sim 1+ O(\epsilon)$, and since
\be
A^{\prime2} e^{2A} H^{-2}\ dr^2 = {({\bf X}\cdot d{\bf X} - T
dT)^2\over {\bf X}^2 - T^2} = H e^{-A}{({\bf X}\cdot d{\bf X} - T
dT)^2\over |{\bf X}^2 - T^2|^{1/2}}\label{disterm}
\ee
(dropping the subscript $(_{p+1})$ on ${\bf X}$ for convenience)
we can obtain (\ref{metric}) by making a small distortion
of the flat metric of the spacetime
in which the vortex-hyperboloid lives.

We now need to consider the analytic extension of the vortex metric
across the horizon. Note that the vortex fields inside the CEH depend
on $r$, which is given implicitly via
\be
{\bf X}_{p+1}^2 - T^2 = H^{-2} e^{2A}
\ee
\ie\ $X$, $A$, and $C$ are constant on spacelike hyperboloids in 
$(T,{\bf X})$-spacetime. In this interpretation, the CEH is the lightcone 
centered on the origin. Therefore, the analytic continuation across the 
CEH would naturally correspond to the interior of the future (or past) 
lightcone of the origin, which means we need the vortex and metric 
functions to depend on $T^2 - {\bf X}^2$ which is a timelike
coordinate. Note that this is \emph{not} the way Nogales and Wang 
tried to continue across the CEH in \cite{W}, they attempted to keep
all fields depending on $r$ and took $r>r_H$. However, the coordinate
transformation in the neighborhood of the CEH shows clearly that
$r>r_H$ corresponds to $z \to z+\pi$ ($z$ being the coordinate along
the 1-dimensional vortex), hence this identifies antipodal points on the
CEH which renders the spacetime singular. It is likely that
this was the singularity they were finding evidence of, for as we
shall see, the CEH is simply a coordinate singularity, and does not
appear to be a future Cauchy horizon, therefore one would not expect
any instabilities there.

Using the de Sitter spacetime as a guide, we therefore look
for a time dependent solution of the Goldstone model (with topological
winding in the scalar field) by replacing (\ref{metric}) with
\be
ds^2 = d\tau^2 - H^{-2} e^{2A(\tau)} \left [ d\xi^2 + \sinh^2
\xi
d\Omega_p^2
\right] - C^2(\tau)\ d\theta^2 \label{extmetric}
\ee
The Einstein equations for this metric, (setting $\Phi=\eta
e^{i\theta}$) give
\bml\bea
{d\over d\tau} \left[{\dot C} e^{(p+1)A}\right] &=&
{2\epsilon e^{(p+1)A} \over C} \label{exthor1}\\
{d\over d\tau} \left[C\,e^{(p+1)A} \,{\dot A}\right]
&=& p H^2 C\,e^{(p-1)A} \label{exthor2}\\
\frac{p}{2}(H^2\,e^{-2A}-{\dot A}^2) -\frac{{\dot A}{\dot
C}}{C}&=&
- \frac{\epsilon }{(p+1)C^2} \label{exthor3}
\eea\eml
which are the same as (\ref{geometry1}-\ref{geometry3}) with
$\epsilon\to
-\epsilon$.

As before, by writing $\rho = -\int e^{-A} dt$, we see that
these equations of motion give the same two dimensional
dynamical system (\ref{xdot},\ref{ydot}), but with the constraint
(\ref{constraint}) now reading
\be\label{extconstr}
{2\epsilon p e^{2A} \over (p+1)C^2} = \left ( x^2 - y^2 - p^2
H^2 \right
)\;.
\ee
Note the minus sign in the definition of $\rho$: this is so that the
event horizon corresponds to the critical point
\be
x = p{dA\over d\rho} + {1\over C} {dC\over d\rho}
= -e^A \left [ p{dA\over d\tau} + {dC\over d\tau} \right]
=-pH
\ee
as $e^A \sim H\tau$ near the horizon.

In other words, whereas the vortex dynamical system corresponded to the
central region in the phase plane in between the two branches of the
invariant hyperboloid, the spacetime `exterior' to the horizon corresponds
to the disconnected regions outside the branches of the invariant
hyperboloid. Indeed, since we know the event horizon corresponds to the
critical point $P_-$ on the left branch of the invariant hyperboloid, we
conclude that the spacetime exterior to the event horizon corresponds
to the left part of this phase plane, $y^2 < x^2 - p^2 H^2$.

The phase plane trajectories on the left of the invariant hyperboloid
generically emerge from the focus (or repeller, depending on the value
of $p$) $Q_-$, and are attracted to the branch of the invariant 
hyperboloid in the upper left hand part of the phase plane, \ie\ 
the Cohen-Kaplan solution which is singular. There is however one trajectory
which terminates on the critical point $P_-$, and this is clearly the
trajectory corresponding to the exterior-horizon solution.
We can therefore read off the late time solution from (\ref{Qpm}) as
\be\label{latet}
ds^2 = d\tau^2 - {p \tau^2\over (p+1)}\left [ d\xi^2 + \sinh^2
\xi
d\Omega_p^2
\right] - {2\epsilon\over (p+1)} \tau^2 d\theta^2 \; .
\ee
This is a late time inhomogeneous open universe cosmology.

Finally, defining
\bml\bea
T &=& H^{-1} e^A \cosh \xi \label{extT}\\
{\bf X}_{p+1} &=& H^{-1} e^A \sinh \xi \ {\bf
n}_{p+1}\label{extX}\\
Z +iY &=& C(\tau) e^{i\theta} \label{extZY}
\eea\eml
now gives
\be
ds^2 = d\tau^2 \left [ H^{-2} {\dot A}^2 e^{2A} - {\dot C}^2 \right ]
- H^{-2} e^{2A} \left [ d\xi^2 + \sinh^2 \xi d\Omega_p^2 \right]
- C^2 d\theta^2 \label{approxext}
\ee
Using the phase plane trajectory of the solution, we can see
that
\be
\left [ H^{-2} {\dot A}^2 e^{2A} - {\dot C}^2 \right ]
= {(x-y)^2\over p^2 H^2} -{2\epsilon
py^2\over(p+1)(x^2-y^2-p^2H^2)}
\ee
can be bounded by
\be
\left ( 1 - {1\over\sqrt{p(p+1)}} \right )^2 - O(\epsilon) <
\left [ H^{-2} {\dot A}^2 e^{2A} - {\dot C}^2 \right ]
< {p+1\over p}
\ee
hence (\ref{approxext}) is indeed a good approximation to the
true exterior metric (\ref{extmetric}).

\begin{figure}[ht]
\begin{center}\epsfxsize=8cm %\epsfysize=4.5cm 
\epsfbox{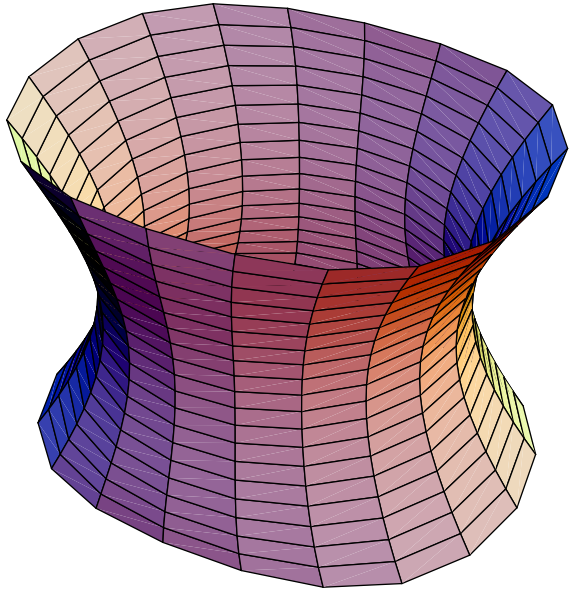}
\end{center}
\caption{The deformed hyperboloid of
the vortex spacetime. The left and right side are close to a genuine
hyperboloid, and correspond to the region inside the CEH. The
`front' and `back' however are flattened, and correspond to the hyperboloid
straightening out to asymptote the flat spacetime outside the CEH.}
\label{fig:vhype}
\end{figure}

Note however that as a surface in Minkowski \real$^{p+4}$, the vortex
departs from the deformed de Sitter-type hyperboloid form that
it had within the event horizon, as
(\ref{extT}-\ref{extZY}) give
\be
2\epsilon (T^2 - {\bf X}^2_{p+1}) - p(Y^2+Z^2) = {2\epsilon
e^{2A}\over H^2} -
pC^2
\simeq 0
\ee
at late times, which is a timelike hyperboloid, and hence spacetime
is asymptotically flat here as required by consistency with (\ref{latet}). 
A sketch of the vortex hyperboloid in
\real$^{p+4}$ is shown in figure \ref{fig:vhype}.
%%%%%%%%%%%%%%
% FIGURE 2
%%%%%%%%%%%%%%

To summarize: we have shown that the vortex (like the domain wall)
compactifies space on a scale roughly of order $O(H^{-1})$.
A $t=$constant section has the topology of a $(p+1)$-sphere, in which
the vortex could be thought of as sitting on a great circle, 
however, as the $S^{p+1}$ is deformed, this picture should be treated
with caution. Nonetheless, we can give the vortex the interpretation
of an accelerating ring (or $p$-sphere) which, like the domain wall,
contracts in from infinity then re-expands out. This picture is most
easily seen by suppressing the $\theta$-direction and using the transformations
(\ref{ring}) to $(T, {\bf X}_{p+1})$ coordinates in which the vortex
follow the trajectory ${\bf X}^2 - T^2 = H^{-2}$.

\section{Strongly gravitating vortices}\label{sect:eps}

An interesting question we can ask is what happens for more
strongly gravitating vortices, \ie\ for $\epsilon\simeq1$. As
$\epsilon$ becomes larger, the spacetime reacts more and more strongly
to the vortex, and the field theory solution departs more and more
from the flat space vortex. Meanwhile, $H$ is increasing, and the
CEH correspondingly moves inwards rather rapidly
leaving less distance outside the vortex core. At some point it 
may well be that the CEH would like to be \emph{inside} the
vortex core. If this occurs, then the vortex fields are essentially
in their false vacuum state and one might expect that the `vortex'
as such disappears and the solution becomes de Sitter, the false vacuum
energy of the unbroken vacuum providing the cosmological constant.
This is in fact the set-up of the topological inflation model of
Linde and Vilenkin \cite{LV}.

In the notionally similar case of the domain wall, it is known that
in the absence of gravitational back reaction, a kink/antikink pair
can be solved exactly on a compact $S^1$ in terms of elliptic 
functions \cite{AI}. It turns out that there is a critical radius
for the circle below which the kink/antikink pair cannot exist and
the only solution is either the false vacuum, or one of the true vacuum
choices. Once gravitational back reaction is correctly taken account
of \cite{BCG} this qualitative result also holds. The gravitational
back reaction of the wall introduces a compactification of space at
a scale of the inverse gravitational coupling ($\epsilon$). As $\epsilon$
grows, this compactification radius shrinks and at some critical
gravitational strength, there is no longer enough space for the kink
to exist and the only possible solution is the false vacuum (which gives
a de Sitter spacetime) or the true vacuum.

Here for the global vortex $p$-brane we no longer have explicit 
solutions\footnote{Local vortices on compact spaces have been studied
\cite{NM}, however to our knowledge, global vortices on spheres have not.},
moreover, $H$ is determined in a rather indirect way, therefore we do not
have as direct an intuition on what happens as $\epsilon$ increases, 
nonetheless, in a manner similar to the arguments of \cite{BCG}, we
can in fact give analytic arguments as to the typical value of
$\epsilon$ at which the vortex solution will cease to exist and the only
possibilities will be the true vacuum or topological inflation.

There are two ways of limiting the value of $\epsilon$ at which the 
vortex ceases to exist. Note that the false vacuum de Sitter (FVDS)
solution, $X\equiv1$ and metric (\ref{deSmet}) with $H^2 = \epsilon/
2(p+1)(p+2)$, is always a solution to the equations of motion (\ref{geometry}).
The first method, which gives a lower bound to $\epsilon_c$, is to
ask at what value of $\epsilon$ this FVDS solution becomes unstable to
vortex formation.  Clearly, if the FVDS solution is unstable to vortex
formation, we are below the critical value of $\epsilon$. The second
method looks at the vortex field equation (\ref{vortexeqn}). If a vortex
spacetime exists, then there will be a solution to this equation (along
with the accompanying $A$ and $C$) which starts with $X=0$ at $r=0$, and
goes to $X=X_H$ at $r=r_H$, with $X'(r_H)=0$ required by nonsingularity
of the spacetime. This places some demands on the behaviour of the function
$X$ which cannot be satisfied if $\epsilon$ is too large. This obviously
gives an upper bound on $\epsilon_c$.

For the first method we need to analyze the perturbation equations
of the FVDS solution:
\be
\Box \xi + {\xi\over C^2} - {\xi\over2} = 0
\ee
It is most transparent to use planar coordinates in the vortex worldvolume
in (\ref{metric}), \ie\ $H^{-2} g_{\mu\nu} dx^\mu dx^\nu 
= dt^2 - e^{2Ht} dx^2_p $ 
in which this perturbation equation takes the form
\be
\left [\ddot{\xi} + pH \dot{\xi} \right] \sec^2 Hr
- \xi'' - \left ( H\cot Hr - (p+1) H \tan Hr \right) \xi' 
+ (H^2 \csc^2 Hr - {\half} ) \xi = 0
\ee
This has solution
\be
\xi = e^{\nu Ht} \sin Hr \cos^\nu Hr
\ee
where
\be
\nu = -{(p+4)\over 2} \pm {1\over 2} \sqrt{ (p+2)^2 + {4\over \epsilon}
(p+1)(p+2)}
\ee
For an instability, we require a perturbation which grows with time, \ie\
$\nu > 0$. Clearly this requires
\be
\epsilon < {(p+1)(p+2)\over(p+3)} \leq \epsilon_c
\ee
This gives the lower bound for the critical value $\epsilon_c$.

For the second method, we now try to look for a vortex-type solution,
with $X'(0)>0$ and $X'$ tending monotonically to zero at the horizon
(we can show this from the equations of motion, but let us simply
state it as a property a vortex solution would display). 
Then, an examination of the equations of motion
(\ref{geometry}) shows that $e^A$ must be monotonically decreasing
from $r=0$ to $r_H$, and $X'(r_H)=0$. Meanwhile, the value of $H$ must
be less than the pure FVDS solution, $H^2 \leq \epsilon/2(p+1)(p+2)$.

Taylor expanding the functions around $r=0$ ($X = x_1r + x_3r^3/6$ etc.) 
and using (\ref{geometry}) gives
\bml\label{taylor}\bea
(p+1)a_2 &=& -{\epsilon\over 4} + {p(p+1)H^2 \over2} < -{\epsilon\over2(p+2)}\\
c_3 &=& -2\epsilon x_1^2 - {\epsilon\over 2(p+1)} - (p+1)a_2\\
x_3 &=& {x_1\over4} \left ( 4\epsilon x_1^2 + {\epsilon\over (p+1)}
- {3\over2} - (p+1)a_2 \right ) 
\eea\eml
We can therefore deduce that if $\epsilon > 3(p+1)/2$, $x_3>0$ and 
$2x_3+c_3x_1>0$. This means that in a neighborhood of the origin,
$X''$ and $F = X'-X/C$ are positive. 

Now let us examine what happens as we move towards the horizon. 
Since $F(r_H) <0$, $F$ must have a zero on $(0,r_H)$ with 
$F'<0$ at this point.  However,
\be\label{Fprime}
F' = \left ( {C\over2} - {C'\over C} - {1\over C} \right ) F
+ {X^3\over2} + X' \left [ -(p+1)A' - {\half C} \right ]
> X' G \Big |_{F=0}
\ee
where $G = \left [ -(p+1)A' - {\half C} \right ]$.
However, from (\ref{taylor}) we see that
$G>0$ near the origin, and indeed
\be
G' = (p+1) A^{\prime2} +
\epsilon F \left ( X' +{X\over C} \right ) + {C'\over C}G 
\ee
is clearly positive while $F$ and $G$ are both positive.
Thus $G>0$ while $F>0$, hence $F$ cannot have a zero from
(\ref{Fprime}).

We conclude that for $\epsilon > 3(p+1)/2$ we cannot have a
vortex-type solution. The critical value of $\epsilon$ therefore lies
in the range
\be
{(p+1)(p+2)\over (p+3)}\leq \epsilon_c \leq {3(p+1)\over2}
\ee
Since $p\geq1$, we see that the value of $\epsilon$ for which
the vortex solution ceases to exist is actually quite high,
$\epsilon_c \geq {3\over2}$ and grows linearly with $p$.
We expect that the CEH in these cases actually occurs for
quite low values of $X_H$.

\section{Discussion}

Since we have shown how the vortex compactifies space on a scale of 
order $H^{-1}$, a natural question to ask is how this feeds back into
a possible braneworld type of resolution of the hierarchy problem.
Briefly, the braneworld picture (pioneered by Rubakov, Shaposhnikov and
Akama \cite{BWP}) imagines that our universe is simply a defect or
submanifold embedded within a higher dimensional manifold. We are confined
to live on this submanifold, but gravity can propagate throughout the
full dimensionality of spacetime. Such models can either have a Kaluza-Klein
type of picture of gravity \cite{ADD}, or can limit long range effects
of the extra dimensions via warped compactifications (e.g.\ \cite{RS}).
The way such models provide an explanation of the hierarchy between
particle and gravitational interaction is via the geometrical volume of
the transverse space (or a factor similar to this for the warped
compactifications).

The hierarchy generation by a vortex compactification was first explored
by Cohen and Kaplan \cite{CK2} for their singular exact solution. In fact,
if we suppose there is no bulk cosmological constant (as CK did) but that
we have our worldvolume Hubble expansion, then we can quite 
straightforwardly derive the hierarchy factor in our case, for writing
the metric in the form 
\be
ds^2 = e^{2A} {\tilde g}_{\mu\nu}(x) dx^\mu dx^\nu - dr^2 - C^2 d\theta^2
\ee
gives
\be
M_n^{p+1} \int \sqrt{g_n} R_n d^n{\rm x} \sim
2\pi M_n^{p+1} \int_0^{r_H} C e^{(p-1)A} dr \int {\tilde R}
\sqrt{{\tilde g}} d^{p+1}x
\ee
which, upon integration of (\ref{geometry2}) from $r=0$ to $r_H$ gives
\be
M_{(p+1)}^{p-1} \simeq {\pi \epsilon M_n^{p+1} \over p(p+1) H^2}
\qquad\Rightarrow\qquad M_{Pl}^2 = {\pi \eta^2 \over 24 H^2}
\ee
for a vortex compactification of six to four dimensions. 
Because of the exponential dependence of $H$ on $\epsilon$, it is
not difficult to get a large hierarchy between the six and four
dimensional Planck masses. Putting $p=3$ in (\ref{Hdef}) and
(\ref{hdef}), and recalling that in vortex units $M_n \simeq 
\epsilon ^{-1/(p+1)}$, gives
\be
{1\over \epsilon} - {9\over8} \ln \epsilon =
4.6 \left ( 16 - \ln ({M_n\over1{\rm TeV}}) \right)
\ee
For a six dimensional Planck scale of 10 TeV, this gives the vortex
scale of around 3.6 TeV, for $M_6 \simeq 100$TeV, the vortex scale
is around 37 TeV.

One interesting feature of this type of compactification is that
the vortex worldvolume is not Minkowski flat spacetime, but rather
an inflating de Sitter universe. Since we suspect that our universe
is in fact asymptoting a mild de Sitter universe \cite{PTW}, 
it is tempting to
use this Hubble expansion to provide the $\Omega_\Lambda \simeq 0.7$
that we see today. Unfortunately, 
the Hubble parameter provided by (\ref{Hdef}), while `small'
in vortex units, is simply too large to provide the tiny $\Omega_\Lambda$
we see today. However, if we try to construct an admixture of
the RS-style warped compactification of \cite{RG3} and the Hubble
expansion of the pure vacuum vortex, we can obtain a more general
solution. 

Technically, the addition of a cosmological constant makes
the  dynamical system phase space three dimensional, however, the
CEH remains a critical point, and now {\it two} of the three eigenvectors
are attractive. This means that there is a two parameter family of
solutions approaching this point, therefore given the initial conditions
outside the core of the vortex, we are still guaranteed to be able to
match this trajectory to one terminating on the critical point by 
varying $H$. We have shown in this paper how one can
have a nonsingular solution with no $\Lambda$ but 
with a worldvolume Hubble expansion, $H$. 
In \cite{RG3} it was shown a nonsingular solution was possible with
negative $\Lambda$ but no $H$. Roughly speaking 
we can therefore find a whole family of
solutions with a cosmological constant and Hubble expansion
$H(\Lambda)$ where $H'(\Lambda)>0$. In principle therefore, by simply
de-tuning the flat Randall-Sundrum type compactification of \cite{RG3},
we can construct a global vortex compactification that both solves
the hierarchy problem as well as giving us the required $\Omega_\Lambda$
we see today. Unfortunately, as there is nothing either `natural' or
generic about this de-tuning, this `resolution' of the cosmological constant
problem is only artificial, in that it simply sweeps the problem
under the rug of higher dimensions.

\acknowledgments

We would like to thank Alex Vilenkin for discussions which motivated this
work, and subsequent comments on the manuscript. This work was 
supported by the Royal Society (RG) and FCT (CS) under the project
CERN/FIS/43737/2001.

\end{document}